\documentclass[10pt,letterpaper,conference]{IEEEtran}
\usepackage{graphicx}
\usepackage[letterpaper, left=1in, right=1in, bottom=1in, top=0.75in]{geometry}
\usepackage{array}
\usepackage{layout}
\usepackage{bm}
\usepackage{algorithmic}
\usepackage{amssymb, amsmath}
\usepackage[numbers, square, comma, sort&compress]{natbib}
\usepackage[ruled,lined]{algorithm2e}
\usepackage{epstopdf}
\usepackage[keeplastbox]{flushend}
\usepackage{subcaption}
\usepackage{siunitx}
\usepackage{url}
\usepackage{xcolor}
\usepackage[english]{babel}
\usepackage{multicol}
\usepackage{color}
\usepackage{acronym}
\usepackage{mathtools}
\usepackage[verbose]{wrapfig}
\usepackage[normalem]{ulem}
\usepackage [autostyle, english = american]{csquotes}
\MakeOuterQuote{"}
\usepackage{caption} 
   \usepackage[belowskip=2pt, aboveskip=-13pt] {caption}
\newcommand{\eq}[1]{(\ref{#1})}

\newcommand{\vmtimet}{{\rm I}(t)}

\acrodef{MEC}[MEC]{Multi-Access Edge Computing}
\acrodef{EPC}[EPC]{Evolved Packet Core}
\acrodef{BS}[BS]{Base Station}
\acrodef{ML}[ML]{Machine Learning}
\acrodef{TIM}[TIM]{Telecom Italia Mobile}
\acrodef{EB}[EB]{Energy Buffer}
\acrodef{EH}[EH]{Energy Harvesting}
\acrodef{QoS}[QoS]{Quality of Service}
\acrodef{MN}[MN]{Mobile Network}
\acrodef{API}[API]{Application Programmable Interfaces}
\acrodef{App}[App]{Application}
\acrodef{ES}[ES]{Energy Saving}
\acrodef{NOES}[NOES]{NO Energy Saving}
\acrodef{VM}[VM]{Virtual Machine}
\acrodef{VUE}[VUE]{Virtual User Equipment} 
\acrodef{RNIS}[RNIS]{Radio Network Information Services} 
\acrodef{LOC}[LOC]{User Location Services} 
\acrodef{EM}[EM]{Energy Manager}
\acrodef{ENAAM}[ENAAM]{ENergy Aware and Adaptive Management}
\acrodef{LLC}[LLC]{Limited Lookahead Controller}
\acrodef{RAN}[RAN]{Radio Access Network}
\acrodef{LSTM}[LSTM]{Long Short-Term Memory}
\acrodef{NFV}[NFV]{Network Function Virtualization}
\acrodef{VNF}[VNF]{Virtualized Network Function}

\begin{document}



\title{\ Online Resource Management in Energy Harvesting BS Sites through Prediction and Soft-Scaling of Computing Resources}


\author{\IEEEauthorblockN {Thembelihle Dlamini\IEEEauthorrefmark{1}\IEEEauthorrefmark{2}, \'Angel Fern\'andez Gamb\'in\IEEEauthorrefmark{1}, Daniele Munaretto\IEEEauthorrefmark{2}, Michele Rossi\IEEEauthorrefmark{1}}
	\IEEEauthorblockA {\IEEEauthorrefmark{1}Department of Information Engineering, University of Padova, Padova, Italy}
	\IEEEauthorblockA {\IEEEauthorrefmark{2}Athonet, Bolzano Vicentino, Vicenza, Italy}
	\{dlamini, afgambin, rossi\}@dei.unipd.it, daniele.munaretto@athonet.com \vspace{-0.4cm}
}

\maketitle
\thispagestyle{plain}
\pagestyle{plain}

\begin{abstract}
Multi-Access Edge Computing (MEC) is a paradigm for handling delay sensitive services that require \mbox{ultra-low} latency at the access network. With it, computing and communications are performed within one \ac{BS} site, where the computation resources are in the form of \acp{VM} (computer emulators) in the MEC server. MEC and \ac{EH} \acp{BS}, i.e., BSs equipped with \ac{EH} equipments, are foreseen as a key towards next generation mobile networks. In fact, \ac{EH} systems are expected to decrease the energy drained from the electricity grid and facilitate the deployment of \acp{BS} in remote places, extending network coverage and making energy \mbox{self-sufficiency} possible in remote/rural sites. In this paper, we propose an online optimization algorithm called \ac{ENAAM}, for managing remote \ac{BS} sites through foresighted control policies exploiting \mbox{(short-term)} traffic load and harvested energy forecasts. Our numerical results reveal that ENAAM achieves energy savings with respect to the case where no energy management is applied, ranging from $56\%$ to $66\%$ through the scaling of computing resources, and keeps the server utilization factor between $30\%$ and $96\%$ over time (with an average of $75\%$). Notable benefits are also found against heuristic energy management techniques.
\end{abstract}


\IEEEpeerreviewmaketitle

\section{Introduction}


\ac{MEC}~\cite{etsimec} ({\it formerly} known as Mobile Edge Computing) has recently emerged as a key solution to process workloads at the network edge, i.e., at the \acp{BS}, while passing the less~\mbox{time-constraint} workloads to the remote cloud. This network design paradigm is based on \ac{NFV}, where mobile network functions (NFs) that formerly existed in the \ac{EPC} are moved to the network edge, such as user's services, which are deployed on local cloud platforms located close to the \acp{BS}. In addition, the 5G \acp{MN} carbon footprint can be minimized through the use of \ac{EH} elements by empowering \acp{BS} with green energy, thus reducing their dependence on the power grid~\cite{ehbs2015}.

The integration of \ac{MEC} and \ac{EH} \acp{BS} can help extend network coverage to areas where the electrical infrastructure cannot reach, or assist during the case of a natural disaster scenario as the network can work in isolation assuming the presence of the \ac{EPC} application in the server, where the conventional electricity grid may become unavailable. Also, it avails computation and storage facilities closer to mobile users, even in remote/rural areas. This will overcome the limitations of current \acp{RAN}, i.e., the lack of computation power and the \mbox{always-on} design approach, as NFV allows the scaling down of some BS functions at low traffic periods or when battery levels are low. However, the integration of \ac{MEC} and \ac{EH} base station systems brings about new challenges related to energy consumption, and resource scheduling. Among other things, quantifying the energy consumed by each running \ac{VM} is a challenge, yet \ac{VM} power metering is key to power consumption minimization in softwarized clouds~\cite{power_metering}.

In~\cite{oh2011toward}~\citep{superbowl}, \acp{ES} towards BSs have been studied to minimize the \ac{BS} power consumption by enabling sleep modes at low traffic load periods. For instance, if a BS has not harvested sufficient energy, its transmission power can be tuned to be in proportion to the energy in its local energy storage and, for low traffic load periods, some of the BS functions can be deactivated. \acp{ES} within the virtualized computing platform are also of great importance. It is known that the power drawn by the server consists of an {\it idle} component and a {\it dynamic} component, which is the power consumed by the physical resources when working on behalf of some \acp{VM}. In~\cite{virttech}, it is shown that power consumption increases with a growth in the number of virtual entities (e.g., \acp{VM}) that are allocated to the physical core, and in~\cite{eempirical}, it is further experimentally shown that increasing the number of VMs also increases power consumption in virtualized platforms, when taking into account the CPU usage only. From the obtained results~\cite{virttech}~\cite{eempirical}, the authors observed that the locus of energy consumption for component of \ac{VNF} is the \ac{VM} instance where the VNF is instantiated/executed. Thus, reducing the number of running \acp{VM} at each time instance, i.e., \ac{VM} \mbox{soft-scaling}, together with BS sleep modes can yield the required energy savings.

Bringing computing and storage services on the \ac{BS} for offloading some workloads requires special attention, as resources are limited at the edge. \mbox{Control-theoretic} and \ac{ML} methods for resource management at the edge have been successfully applied to various problems, e.g., task scheduling, bandwidth allocation, network management policies, etc. In~\cite{llcprediction}, the authors presented a generic online control framework for resource management in switching hybrid systems, where the system's control inputs are finite. The relevant parameters of the operating environment, e.g., workload arrival, are estimated and then used by the system to forecast future behavior over a \mbox{look-ahead} horizon. From this, the controller optimizes the predicted system behavior following the specified \ac{QoS} through the selection of the system control inputs. In~\cite{chung1992limited}, a supervisory online control scheme based on \ac{LLC} policies is presented. The authors in~\cite{xu2016online} presents a reinforcement \mbox{learning-based} resource management algorithm to incorporate renewable energy into a \ac{MEC} platform. At the beginning of the time slot the servers are consolidated, i.e., the number of turned on physical servers are minimized, using the learned optimal policy for dynamic workload offloading and the autoscaling (or right-sizing). Our work differs from~\cite{xu2016online}, as we minimize the number of active \acp{VM} instead of server consolidation, and also we use a forecasting method instead of only relying on the available current information for decision making.


\noindent \textbf{Paper contributions:} here, we consider the aforementioned scenario, where the \ac{BS} is equipped with \ac{EH} hardware and computation capabilities, i.e., a solar panel or wind turbine for \ac{EH}, an energy storage unit termed \ac{EB}, and a \ac{MEC} server \mbox{co-located} with the BS. The presence of \ac{EB} and \ac{EH} systems promotes energy \mbox{self-sustainability} and network coverage extension. Motivated by the potential of \ac{EH} and \ac{MEC},
\textbf{1)} we estimate the \mbox{short-term} future traffic load and harvested energy in \acp{BS}, by using {\it Recurrent Neural Networks} (RNNs~\cite{machine_learning_tut}), specifically a \ac{LSTM} network, coupled with forecasting knowledge from~\cite{forecasting},
and
\textbf{2)}  we develop an online algorithm for edge network management based on control theory. The main goal is to enable \ac{ES} strategies within the remote site through \ac{BS} sleep modes and \ac{VM} \mbox{soft-scaling}, following the energy efficiency requirements of a virtualized infrastructure from~\cite{eerequirements}. The proposed management algorithm is called ENergy Aware and Adaptive Management (ENAAM) and is hosted in the \ac{MEC} server, i.e., \ac{ENAAM} \ac{App}. The \ac{ENAAM} application considers the future BS loads, onsite green energy in the \ac{EB} and then provisions edge network resources based on the learned information, i.e., \ac{ES} decisions are made in a \mbox{forward-looking} fashion.


The proposed optimization strategy reads to a considerable reduction in the energy consumed by edge computing and communication facilities, enabling mobile services to \mbox{off-grid} sites under limited energy budget and predicted traffic loads.

The rest of the paper is structured as follows: the system model is presented in Section~\ref{sec:sys}. In Section~\ref{sec:prob}, we detail the optimization problem and the proposed \ac{ENAAM} online algorithm. In Section~\ref{sec:results}, we evaluate our online edge network management algorithm and lastly, we conclude our work in Section~\ref{sec:concl}.

\begin{figure}[t]
\centering
\includegraphics[width =0.48\textwidth]{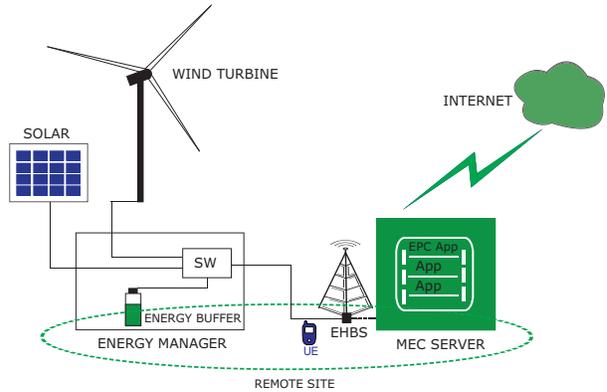}
\caption{\ac{EH} \ac{BS} \mbox{co-located} with a \ac{MEC} server. The switch (SW) is responsible for selecting the appropriate renewable energy source to power the \ac{BS} site.}
\label{fig:mecscenario}
\end{figure}

\section{System Model}
\label{sec:sys}

The considered scenario is illustrated in Fig.~\ref{fig:mecscenario}. As a major deployment method of \ac{MEC}, we consider a setup where a \ac{BS} is \mbox{co-located} with a virtualized \ac{MEC} server, forming a communication site {\it termed} remote site. Both share the available energy stored in the \ac{EB}. Following the motivation from the introduction, we only consider an {\it offgrid} BS \mbox{co-located} with the MEC server. The MEC server accounts for $M$ virtual machines as total computation resources, and it is \mbox{cache-enabled}, i.e., some contents can be accumulated locally. Radio network and energy level information is reported periodically to the MEC server through \ac{RNIS} and the \ac{EM}, an entity responsible for selecting the appropriate renewable energy source to fulfill the \ac{EB} depending on the weather, and for monitoring the energy levels in the system. Moreover, we consider a \mbox{discrete-time} model, whereby time is discretized as $t = 1,2,\dots$, and time slots have a constant duration.
For data communication from the remote site to the remote cloud, the system uses a microwave backhaul, as fast \mbox{roll-out} over large distances makes microwave an ideal rural/remote backhaul solution~\cite{microwave_eri}.

\subsection{Traffic Load and Power Consumption}
\label{sub:bsload}

\begin{figure}[t]
	\centering
	\includegraphics[width =\columnwidth]{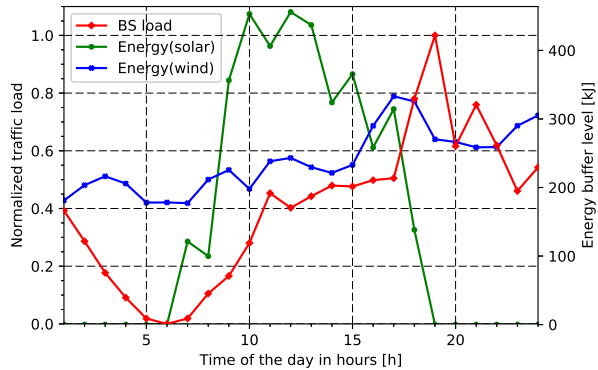}
	\caption{Example traces for harvested solar and wind energy, and normalized traffic load in the BS.}
	\label{fig:harv_load}
\end{figure} 

Traffic load traces have been obtained using real \ac{MN} data from the Big Data Challenge organized by \ac{TIM}~\cite{bigdata2015tim}. The open source dataset is a result of users interaction within the \ac{TIM} MN for the city of Milan during the month of November 2013, whereby each interaction generates a Call Detail Record (CDR) file. The considered TIM dataset refers to standard traffic such as SMS, Calls and Internet browsing, and they are not yet a representative of future applications that require processing at the edge. In this paper, according to~\cite{youtube_pareto}, we assume that $80$\% of the traffic from this dataset requires processing at the edge, whereas the remaining $20$\% pertains to standard, delay tolerant, flows. The daily traffic load profile requiring computation at the BS, ${\rm L}(t)$, see red curve in Fig.~\ref{fig:harv_load}, is obtained by accounting for $80$\% of the aggregated CDR data. The normalized  \ac{BS} load at time slot $t$ is approximated as $\varphi (t) = {\rm L}(t)/ {\rm L}_{\rm max}$, where ${\rm L}_{\rm max}$ represents the maximum load that can be served. Among this load, $\gamma(t) \in [0,1]$ is processed locally and the rest $\Gamma(t) = \varphi (t) - \gamma(t) \in [0,1]$ is handled by the remote cloud. Moreover, a low traffic threshold $L_{\rm low}$ is defined to be used in the ENAAM algorithm (see subsection~\ref{sec:alg}). Note that $\gamma(t)$ is a decision variable, as we shall see, $\gamma(t)$ is ideally set to $1$ for those time slots where the BS has enough energy, i.e., all the delay sensitive traffic is processed at the edge. $\gamma(t)$ will be set to a smaller value otherwise.

The total energy consumption ([$\SI{} {\kilo\joule}$]) of the remote site is here obtained as the combination of the energy consumed by the \ac{BS} and by the \mbox{co-located} MEC server operating at a frequency $f$ ([$\SI{}{\hertz}$]), with server maximum utilization factor $\gamma(t)$ in time slot $t$. The following model is inspired by~\cite{bspower} and~\cite{Liumigration}, by additionally tuning the BS static energy and the server utilization factor, to scale the server dynamic energy consumption in proportion to the expected load to be processed locally,
\begin{equation}
	\theta(\zeta,\gamma,t) = \zeta(t) \theta_{0} + \theta_{\rm tx}(t) +  \theta_{\rm bh} + \theta_{\rm mec}(\gamma,t) \, ,
	\label{eq:bsconsupt}
\end{equation}
\noindent where $\zeta(t) \in  \{\varepsilon, 1\}$ is the BS switching status indicator ($1$ for {\it active mode} and $\varepsilon$ for {\it power saving mode}), $\theta_{0}$ is a constant value (load independent), representing the operation energy which includes baseband processing, radio frequency power expenditures, etc. The constant $\varepsilon \in (0,1)$ accounts for the fact that the baseband energy consumption can be scaled down as well whenever there is no or little channel activity, into a power saving mode. $\theta_{\rm tx}(t)$ represents the total downlink transmission (load dependent) power from the BS to the served user(s). Since we assume a \mbox{noise-limited} channel and the guarantee of low latency requirements at the edge, to obtain $\theta_{\rm tx}(t)$ we use the downlink transmission model in~\cite{mec_lyapunov} (see Eq.~(4) in that paper). $\theta_{\rm bh}$ is the microwave backhaul transmission energy, which is here assumed to be constant. $\theta_{\rm mec}(\gamma,t)$ is the  computation energy at the server, defined as follows: \mbox{$\theta_{\rm mec} (\gamma, t) = \theta_{\rm idle} + \gamma(t) \theta_{\rm dyn}(t)$},
where $\theta_{\rm idle}(\cdot)$ is the server \mbox{load-independent} operational component, and $\theta_{\rm dyn}(\cdot)$ is the maximum energy amount that is consumed by the server when it operates at full power. Although omitted for the sake of notation compactness, $\theta_{\rm idle}$ and $\theta_{\rm dyn}(\cdot)$ depend on the MEC server computation frequency $f$. Also, $\theta_{\rm dyn}(\cdot)$ is linearly scaled with respect to the load $\gamma(t)$, assuming that computation resources can be tuned. Finally, the number of virtual machines that shall be active in time slot $t$ to serve the offered load is here obtained as $\vmtimet = \texttt{round}(\gamma(t) M)$, where \texttt{round}$(\cdot)$ rounds the argument to the nearest integer.

\subsection{Energy Patterns and Storage}
\label{sub:eebuffer}

The energy buffer is characterized by its maximum energy storage capacity $\beta_{\rm max}$. At the {\it beginning} of each time slot $t$, the EM provides the energy level report to the MEC server application, thus the \ac{EB} level $\beta(t)$ is known, enabling the provision of the required computation resources, i.e., \acp{VM}. Here, a pull transfer mode (e.g., FTP~\cite{filetransfer}) is assumed, where the MEC application pulls the energy report from the EM. The amount of harvested energy $H(t)$ in time slot $t$ for the remote site is obtained from open source solar traces from~\cite{amerinia} (see green curve in Fig.~\ref{fig:harv_load}), and also wind traces from~\cite{belgium} (see blue curve in Fig.~\ref{fig:harv_load}). The datasets are a result of daily environmental records, considering solar panel orientation, measured and forecast wind speed, temperature, wind power, and pressure values. In this paper, $H(t)$ is obtained by first scaling the datasets to fit the \ac{EB} capacity $\beta_{\rm max}$ of $\SI{490} {\kilo\joule}$, and then selecting the wind energy as a source during the solar energy \mbox{off-peak} periods. Thus, the available \ac{EB} level $\beta(t + 1)$ for the {\it offgrid} BS in time slot $t+1$ is calculated as follows: 

\begin{equation}
\beta(t + 1) = \beta(t) + H(t) - \theta(\zeta,\gamma,t) 
\label{eq:offgrid}
\end{equation}

\noindent where $\beta(t)$ is the energy level in the battery at the beginning of time slot $t$ and $\theta(\zeta,\gamma,t)$ is the energy that is used during the time slot for computation and communications, see Eq.~(\ref{eq:bsconsupt}). For decision making in the MEC server, a lower battery threshold is defined, $\beta_{\rm low}$, with $0 < \beta_{\rm low} < \beta_{\rm max}$, to steer how the energy management algorithm provisions the required edge network resources, see Section~\ref{sub:res}. 

\section{Problem Formulation}
\label{sec:prob}

In this section, we formulate an optimization problem to obtain {\it energy savings} through \mbox{short-term} traffic load and harvested energy predictions along with energy management procedures. The optimization problem is defined in section~\ref{sub: opt_prob}, and the remote site management procedures are presented in Section~\ref{sub:res}. 

\subsection{Optimization Problem}
\label{sub: opt_prob}

At the beginning of each time slot $t$, the MEC server receives the energy level report $\beta(t)$ from the EM. 
In this paper, we aim at minimizing the overall energy consumption in the remote site over time, i.e., consumption related to the \ac{BS} and MEC server, by applying BS power saving modes and \ac{VM} \mbox{soft-scaling}, i.e., tuning the number of active virtual machines. To achieve this, we define two cost functions: F1) $\theta(\zeta,\gamma,t)$, which weighs the energy consumption due to transmission (BS) and computation (MEC server); and F2) a quadratic term $(\varphi(t)-\gamma(t))^2$, which accounts for the \ac{QoS} cost. In fact, F1 tends to push the system towards energy efficient solutions, i.e., where $\gamma(t) \to 0$ and $\zeta(t) \to \varepsilon$. Instead, F2 favors solutions where the load is entirely processed by the local MEC server, i.e., where $\gamma(t) \to \varphi(t)$. A weight $\alpha \in [0,1]$, is utilized to balance the two objectives F1 and F2. The corresponding weighted cost function is defined as: 
\begin{equation}
\label{eq:Jfunc}
J(\zeta,\gamma,t) \stackrel{\Delta}{=} \overline{\alpha} \theta(\zeta(t),\gamma(t),t) + \alpha (\varphi(t)-\gamma(t))^2 \, ,
\end{equation}
where $\overline{\alpha} \stackrel{\Delta}{=} 1 - \alpha$. Hence, over time horizon, $t=1,\dots,T$, the following optimization problem is defined:
\begin{eqnarray}
	\label{eq:objt}
	\textbf{P1} & : & \min_{\bm \zeta, \bm \gamma} \sum_{t=1}^T J(\zeta,\gamma,t)  \nonumber \\
	&& \hspace{-1.25cm}\mbox{subject to:} \nonumber \\
	{\rm C1} & : & 0 < \gamma(t) \leq 1, \quad t=1,\dots, T \nonumber \\
	{\rm C2} & : & \zeta(t) \in \{\varepsilon,1\}, \quad t=1,\dots, T \\
	{\rm C3} & : & {\rm I}(t) \geq b, \quad t=1,\dots, T \nonumber \\ 
	{\rm C4} & : & \beta(t) \geq \beta_{\rm low} , \quad t=1,\dots, T \nonumber \, 	
\end{eqnarray}

\noindent where vectors $\bm \zeta$ (switching status) and $\bm \gamma$ (utilization factor) contain the control actions for the considered time horizon $1,2,\dots,T$, i.e., $\bm \zeta = [\zeta(1), \zeta(2), \dots, \zeta(T)]$ and \mbox{$\bm \gamma = [\gamma(1), \gamma(2), \dots, \gamma(T)]$}. Constraint C1 specifies the server utilization factor bounds, C2 specifies the BS operation status, C3 forces the required number of \acp{VM}, ${\rm I}(t)$, to be always greater than or equal to a minimum number \mbox{$b \geq 1$}: the purpose of this is to be always able to handle mission critical communications. C4 makes sure that the \ac{EB} level is always above or equal to a preset threshold $\beta_{\rm low}$, to guarantee energy \mbox{self-sustainability} over time. To solve {\rm P1} in Eq.~\eq{eq:objt}, we leverage the use of \ac{LLC}~\cite{llcprediction}~\cite{chung1992limited} and heuristics. Once P1 is solved, the control action to be applied at time $t$ is $\varsigma(t)\stackrel{\Delta}{=}(\zeta(t),\gamma(t))$.

\subsection{Remote Site Management}
\label{sub:res}

In this subsection, a traffic load and energy harvesting prediction method, and an online management algorithm are proposed to solve the previously stated problem {\rm P1}. In subsection~\ref{predict}, we discuss the machine learning tool used to predict the \mbox{short-term} future traffic loads and harvested energy, and then in subsection~\ref{online_proc}, we solve {\rm P1} by first constructing the \mbox{state-space} behavior of the control system, where online control key concepts are introduced. Finally, the algorithm for managing the remote site is presented in subsection~\ref{sec:alg}.

\subsubsection{Traffic load and energy prediction}
\label{predict}

\ac{ML} techniques constitute a promising solution for network management and energy savings in cellular networks~\cite{machine_learning_tut}\cite{machine_learning_details}. In this work, we consider a time slot duration of one hour and perform time series prediction, i.e., we obtain the \mbox{$\SI{1} {\hour}$-ahead} estimates of $\hat{\rm L}(t+1)$ and $\hat{H}(t+1)$, by using an \ac{LSTM} developed in Python using Keras deep learning libraries (Sequential, Dense, LSTM) where the network has a visible layer with $1$ input, a hidden layer of $4$ \ac{LSTM} blocks or neurons, and an output layer that makes a single value prediction. This type of recurrent neural network uses \mbox{back-propagation} through time and memory blocks for regression~\cite{lstmlearn}. The dataset is split as $67\%$ for training and $33\%$ for testing. The network is trained using $100$ epochs ($2,600$ individual training trials) with batch size of $1$. As for the performance measure of the model, we use the Root Mean Square Error (RMSE). The prediction steps are outlined in Table~\ref{lstm_model}, and Fig.~\ref{fig:bs_load} show the prediction results that will be discussed in Section~\ref{sec:results}.


\begin{table}[tp]
	\caption{LSTM Prediction Model Steps}
	\center
	\resizebox{\columnwidth}{!}{%
		\begin{tabular} {|l|}
			\hline 
			{\bf Modeling steps} \\ 
			\hline
			Step 1: load and normalize the dataset\\
			Step 2: split dataset into training and testing\\
			Step 3: reshape input to be [samples, time steps, features]\\
			Step 4: create and fit the LSTM network\\
			Step 5: make predictions\\
			Step 6: calculate performance measure\\
			\hline 
		\end{tabular}%
	}
	\label{lstm_model}
\end{table}

\subsubsection{Edge system dynamics}
\label{online_proc}

we denote the system state vector at time $t$ by $x(t) = (\vmtimet,\beta(t))$, which contains the number of active VMs, and the EB level. \mbox{$\varsigma(t)=(\zeta(t),\gamma(t))$} is the input vector, i.e., the control action that drives the system behavior at time $t$. The system evolution is described through a \mbox{discrete-time} \mbox{state-space} equation, adopting the \ac{LLC} principles~\cite{llcprediction}~\cite{chung1992limited}:
\begin{equation}
x(t + 1) = \Phi(x(t),\varsigma(t)) \, , 
\end{equation}
\noindent where  $\Phi(\cdot)$ is a behavior model that captures the relationship between $(x(t),\varsigma(t))$, and the next state $x(t + 1)$. Note that this relationship accounts for 1) the amount of energy drained $\theta(\zeta,\gamma,t)$ and the harvested $H(t)$, which together lead to the next buffer level $\beta(t+1)$ through Eq.~\eq{eq:offgrid}, and 2) to the traffic load $L(t)$, from which we compute the offered load $\varphi(t)$, that together with the control $\gamma(t)$ leads to ${\rm I}(t+1)$ (once a control policy is specified). The remote site management ENAAM App, acts as a controller, that finds the best control action vector to the system, iteratively. For each time slot $t$, the best control action $\varsigma^{*}(t)$ is the one minimizing the weighted sum $J(\zeta,\gamma,t)$. This control action amounts to setting the BS radio mode $\zeta^*(t)$, i.e., either active or power saving, and the number of instantiated \acp{VM}, ${\rm I}^*(t)$, which directly follows from $\gamma^*(t)$. 

An observation is in order. State $x(t)$ and control $\varsigma(t)$ are respectively measured and applied at the beginning of time slot $t$, whereas the offered load $L(t)$ and the harvested energy $H(t)$ are accumulated during the time slot and their value becomes known only by the end of it. This means that, being at the beginning of time slot $t$, the system state at the next time slot $t+1$ can only be estimated, which we formally write as:
\begin{equation}
	\hat{x}(t + 1) = \Phi(x(t),\varsigma(t)) \, . 
	\label{eq:state_forecast}
\end{equation}


\noindent\textbf{Controller decision-making:} the controller is obtained by estimating the relevant parameters of the operating environment, that in our case are the BS load $\hat{L}(t)$ and the harvested energy $\hat{H}(t)$, and subsequently using them to forecast the future system behavior through Eq.~\eq{eq:state_forecast} over a \mbox{look-ahead} time horizon of $T$ time slots.  
The control actions are picked by minimizing $J(\zeta,\gamma,t)$, see Eq.~\eq{eq:Jfunc}. At the beginning of each time slot $t$ the following process is iterated: 

\begin{itemize} 
\item Future system states, $\hat{x}(t+k)$, for a prediction horizon of $k = 1, \dots, T$ steps are estimated using Eq.~\eq{eq:Jfunc}. These predictions depend on past inputs and outputs up to time $t$, on the estimated load $\hat{L}(\cdot)$ and energy harvesting $\hat{H}(\cdot)$ processes, and on the control $\varsigma(t+k)$, with $k =  0, \dots, T-1$.
\item The sequence of controls $\{\varsigma(t+k)\}_{k=0}^{T-1}$ is obtained for each step of the prediction horizon by optimizing the weighted cost function $J(\cdot)$.
\item The control $\varsigma^*(t)$ corresponding to the first control action in the sequence with the minimum total cost is the applied control for time $t$ and the other controls $\varsigma^*(t+k)$ with $k = 1, \dots, T-1$ are discarded. 
\item At the beginning of the next time slot $t+1$, the system state $x(t+1)$ becomes known and the previous steps are repeated. 
\end{itemize}

\begin{small}
\begin{algorithm}[t]
\begin{tabular}{l l}
{\bf Input:}  & $x(t)$ (current state) \\
{\bf Output:} & $\varsigma^{*}(t) = (\zeta^*(t),\gamma^*(t))$ \\
01:		& \hspace{-1cm}Initialization of variables\\
		& \hspace{-1cm}${\mathcal S}(t) = \{x(t)\}$, ${\rm Cost}(x(t))=0$ \\
02:		& \hspace{-1cm}{\bf for} $k = 1, \dots, T$ {\bf do}\\
		& \hspace{-1cm}\quad - forecast the load $\hat{\rm L}(t+k-1)$ \\
		&\hspace{-1cm}\quad - forecast the harvested energy\\
		& $\hat{\rm H}(t+k-1)$ \\
		& \hspace{-1cm}\quad - ${\mathcal S}(t+k) = \emptyset$ \\
03:		& \hspace{-1cm}\quad {\bf for all} $x \in {\mathcal S}(t+k-1)$ {\bf do}\\
04:		& \hspace{-1cm}\qquad {\bf for all} $\varsigma = (\zeta,\gamma) \in {\mathcal A}(t+k-1)$ {\bf do}\\
05:		& \hspace{-1.1cm}\quad\quad\quad $\hat{x}(t+k) = \Phi(x(t+k-1),\varsigma)$\\
06:		& \hspace{-1.1cm}\quad\quad\quad ${\rm Cost}(\hat{x}(t+k)) =  J(\zeta, \gamma, t+k-1)$\\
		& \hspace{1.25cm} $ + {\rm Cost}(x(t+k-1),\varsigma)$\\
07:		& \hspace{-1.1cm}\quad\quad\quad ${\mathcal S}(t+k) = {\mathcal S}(t+k) \cup \{\hat{x}(t+k)\}$\\
		& \hspace{-1cm}\qquad {\bf end for}\\
		& \hspace{-1cm}\quad {\bf end for}\\
		& \hspace{-1cm}{\bf end for}\\
08:		& \hspace{-1cm}{\bf Find $\hat{x}_{\min} = {\rm argmin}_{\hat{x} \in {\mathcal S}(t+T)} {\rm Cost}(\hat{x}) $}\\
09:		& \hspace{-1cm}{$\varsigma^{*}(t) :=$ control leading from $x(t)$ to $\hat{x}_{\min}$}\\
10:		& \hspace{-1cm}{\bf Return $\varsigma^{*}(t)$}
\end{tabular}
\caption{ENAAM}
\label{algo:enaam}
\end{algorithm}
\end{small}

\subsubsection{The ENAAM algorithm}
\label{sec:alg}

Let $t$ be the current time. $\hat{\rm L}(t+k-1)$ is the forecast load in slot $t+k-1$, with $k=1,\dots,T$, i.e., over the prediction horizon. For the control to be feasible, we have $\hat{\varphi}(t+k-1) = \hat{\rm L}(t+k-1) / L_{\max}$ and $\underline{\gamma} \leq \gamma \leq \hat{\varphi}(t+k-1)$, where $\underline{\gamma}$ is the smallest $\gamma$ such that $\texttt{round}(\underline{\gamma} M) = b$. For the buffer state, we heuristically set $\zeta(t+k-1) = \varepsilon$ if $\beta(t+k-1) < \beta_{\rm low}$ {\it or} $L(t+k-1) < L_{\rm low}$, and $\zeta(t+k-1) = 1$ otherwise ($\beta_{\rm low}$ and $L_{\rm low}$ are preset thresholds). For slot $t+k-1$, the feasibility set $\mathcal A(t+k-1)$ contains the control pairs $(\zeta,\gamma)$ that obey these relations.

The algorithm is specified in Alg.~\ref{algo:enaam} as it uses the technique in~\cite{llcprediction}: the search starts (line 01) from the system state at time $t$, $x(t)$, and continues in a \mbox{breadth-first} fashion, building a tree of all possible future states up to the prediction depth $T$. A cost is initialized to zero (line 01) and is accumulated as the algorithm travels through the tree (line 06), accounting for predictions, past outputs and controls. The set of states reached at every prediction depth $t+k$ is referred to as $\mathcal S(t+k)$. For every prediction depth $t+k$, the search continues from the set of states $\mathcal S(t+k-1)$ reached at the previous step $t+k-1$ (line 03), exploring all feasible controls (line 04), obtaining the next system state from Eq.~\eq{eq:state_forecast} (line 05), updating the accumulated cost as the result of the previous accumulated cost, plus the cost associated with the current step (line 06), and updating the set of states reached at step $t+k$ (line 07). When the exploration finishes, the action at time $t$ that leads to the best final accumulated cost, at time $t+T$, is selected as the optimal control $\varsigma^*(t)$ (lines 08, 09, 10). Finally, for line 04, we note that $\gamma$ belongs to the continuous set $[\underline{\gamma}, \hat{\varphi}(t+k-1)]$. To implement this search, we quantized this interval into a number of equally spaced points, obtaining a search over a finite set of controls.  

\section{Performance Evaluation}
\label{sec:results}

In this section we show some selected numerical results for the scenario of Section~\ref{sec:sys}. The parameters that were used for the simulations are listed in Table~\ref{tab_opt}.

\noindent \textbf{Simulation Setup:} we consider one offgrid \ac{BS} co-located with a MEC server within a coverage area of $\SI{40} {\meter}$. In addition, we use a virtualized server with specifications from~\cite{specs_online} for a VMware ESXi \mbox{5.1-ProLiant} DL380 Gen8 that operates at $f = 1,600$~MHz. Our time slot duration is set to $\SI{1} {\hour}$ and the time horizon to $T=2$. For our simulations, Python is used as the programming language.

\begin{figure}[t]
	\centering
	\includegraphics[width =\columnwidth]{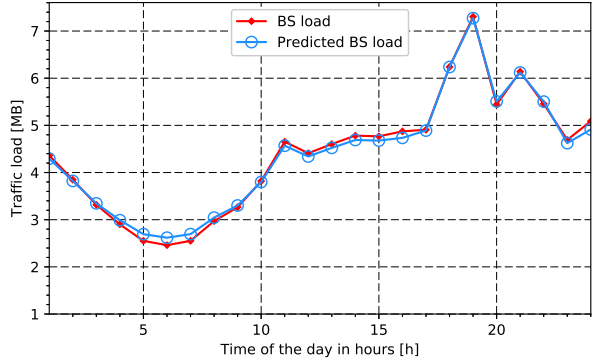}
	\caption{One-step ahead predicted BS load (LSTM).}
	\label{fig:bs_load}
\end{figure}

\noindent \textbf{Numerical Results:} some example prediction results are shown in Fig.~\ref{fig:bs_load} for the traffic load, reaching an RMSE performance of $0.42$~MB. Quite good accuracies are also obtained for the prediction of the harvested energy (RMSE of $0.38$~kJ over a range of about $450$~kJ). The measured accuracy is deemed good enough for the proposed optimization.  

\begin{table}[t]
	\caption{System Parameters.}
	\center
	\begin{tabular} {|l| l|l|}
		\hline 
		{\bf Parameter} & {\bf Value} \\ 
		\hline
		Low traffic threshold, $L_{\rm low}$ & $4$~MB \\ 
		Maximum load, $L_{\rm max}$ & $15$~MB \\
		Operating power, $\theta_{0}$ & $\SI{10.6} {\watt}$\\
		Microwave backhaul power, $\theta_{\rm bh}$ & $\SI{50} {\watt}$\\
		Maximum number of \acp{VM}, $M$ &  $27$\\
		Minimum number of \acp{VM}, $b$ & $3$ \\
		Idle power, $\theta_{\rm idle}$ & $\SI{30} {\watt}$\\
		Dynamic maximum power, $\theta_{\rm dyn} (t)$ & $\SI{472.3} {\watt}$\\
		Energy storage capacity, $\beta_{\rm max}$ & $\SI{490} {\kilo\joule}$\\
		Lower energy threshold, $\beta_{\rm low}$  & $30$\% of $\beta_{\rm max}$\\
		Maximum number of served users & $50$\\
		\hline 
	\end{tabular}
	\label{tab_opt}
\end{table}

\begin{figure}[t]
	\centering
	\includegraphics[width =\columnwidth]{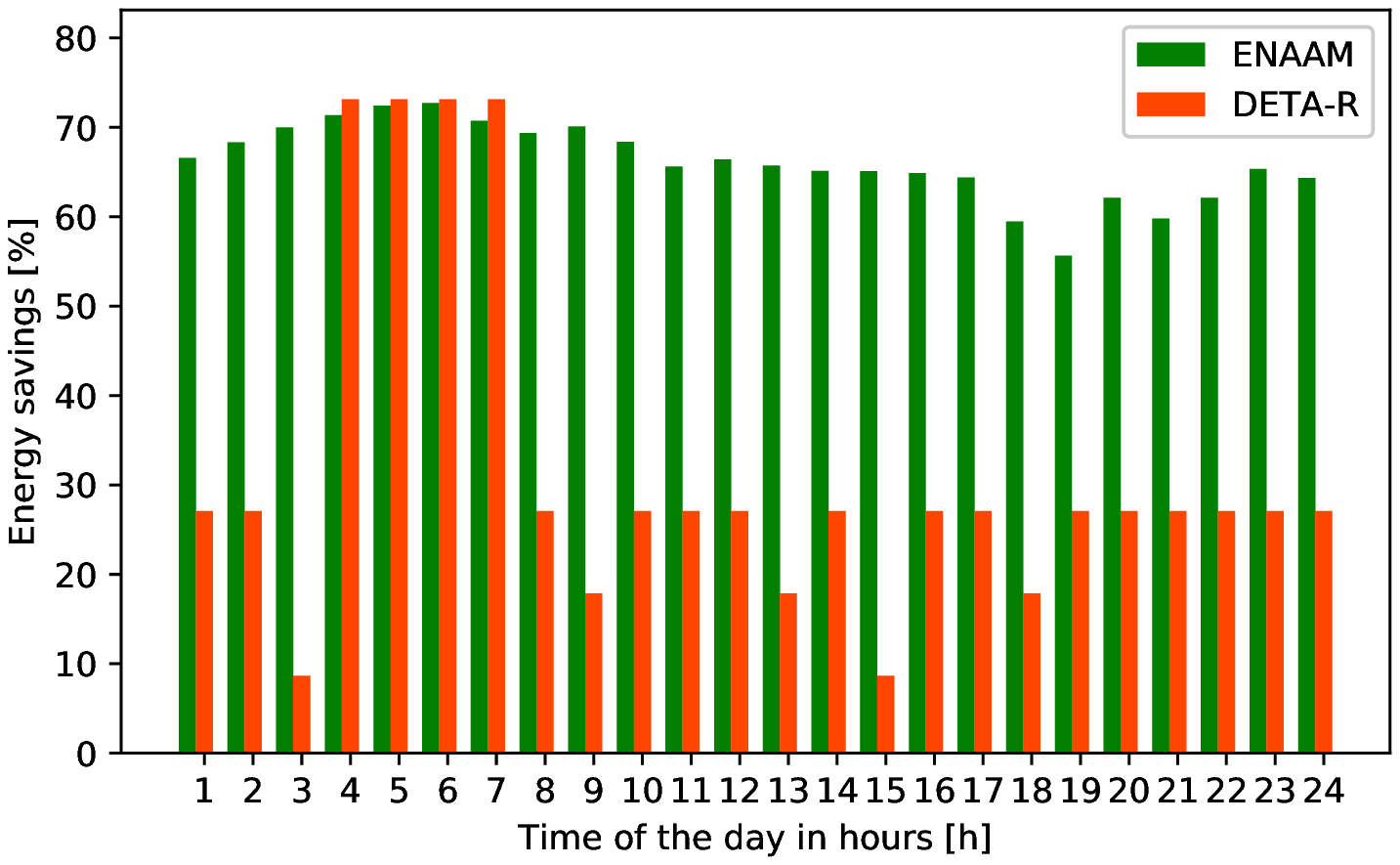}
	\caption{Hourly energy savings for $\alpha = 0$.}
	\label{fig:savings_0}
\end{figure} 

\begin{figure}[t]
	\centering
	\includegraphics[width =\columnwidth]{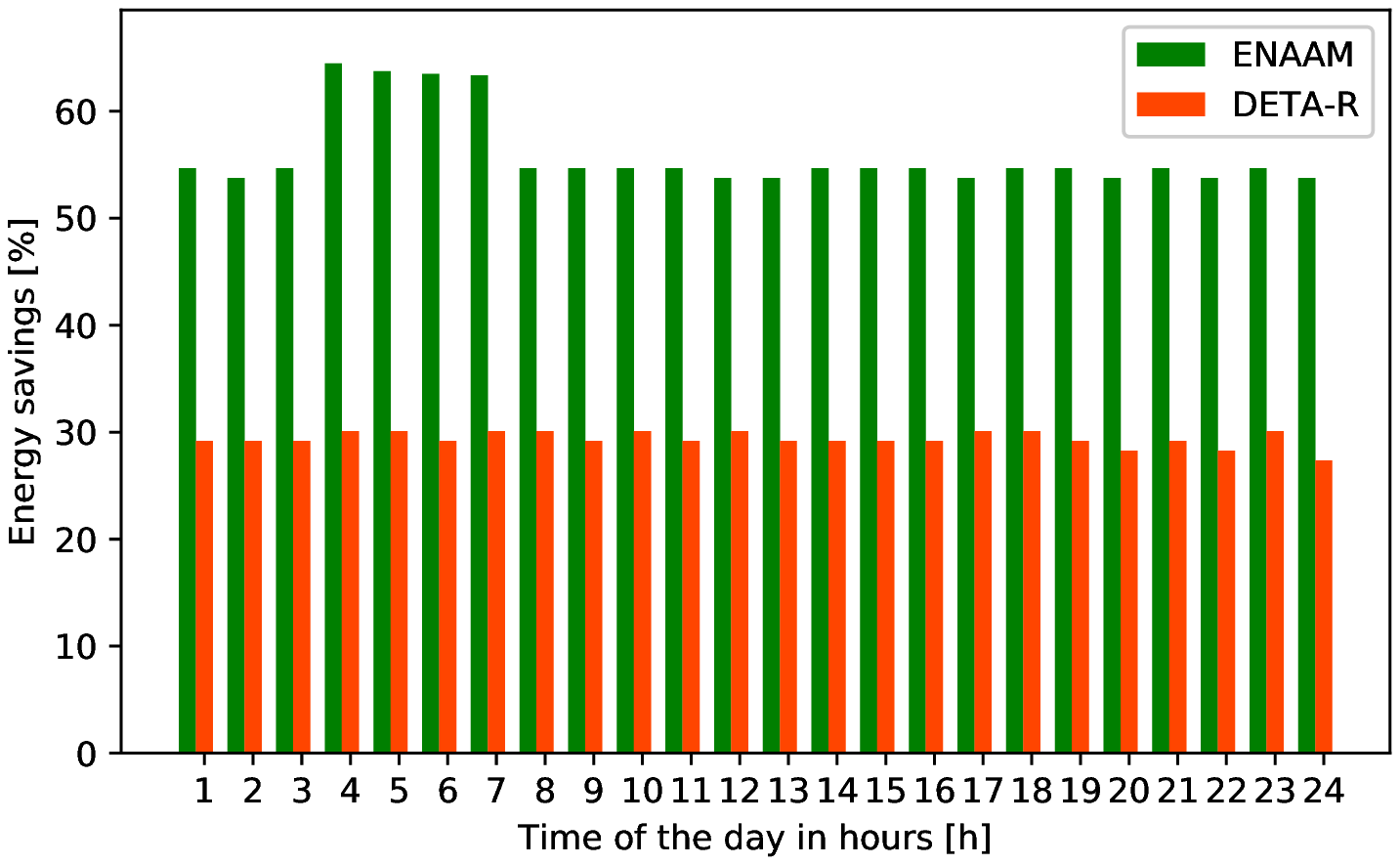}
	\caption{Hourly energy savings for $\alpha = 0.5$.}
	\label{fig:savings_0.5}
\end{figure} 

\begin{figure}[t]
	\centering
	\includegraphics[width =\columnwidth]{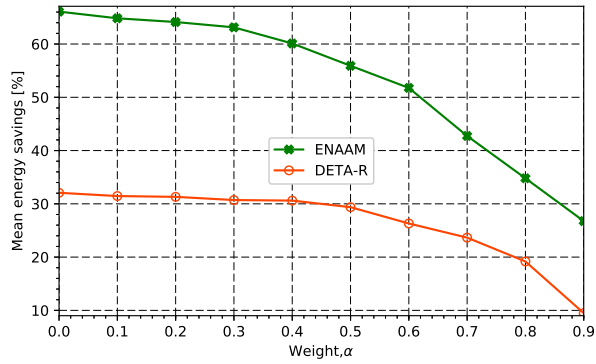}
	\caption{Energy savings {\it vs} the optimization weight $\alpha$.}
	\label{fig:alpha_val}
\end{figure}

Figs.~\ref{fig:savings_0} and~\ref{fig:savings_0.5} show the energy savings achieved over time for $\alpha = 0$ and $\alpha = 0.5$ respectively, when \mbox{on-demand} and \mbox{energy-aware} edge resource provisioning is enabled (i.e., BS sleep modes and VM soft-scaling), in comparison with the case where they are not applied. Our remote site management algorithm (\ac{ENAAM}) is benchmarked with another one that heuristically selects the amount of traffic that is to be processed locally, $\gamma(t)$, depending on the expected load behavior. It is named Dynamic and \mbox{Energy-Traffic-Aware} algorithm with Random behavior \mbox{({DETA-R})}. Both ENAAM and \mbox{DETA-R} are aware of the predictions in the future time slots, see Section~\ref{predict}, however, \mbox{DETA-R} provisions edge resources using a heuristic scheme. \mbox{DETA-R} heuristic works as follows:  if the expected load difference is \mbox{$\hat{\rm L}(t+1) - \hat{\rm L}(t) >0$}, then $\gamma(t)$ in the current time slot $t$ is randomly selected in the range $[0.6, 1]$, otherwise, it is picked evenly at random in the range $(0,0.6)$.

When $\alpha = 0$, Fig.~\ref{fig:savings_0} shows energy savings of $66\%$ on average when ENAAM scheme is applied, while \mbox{DETA-R} achieves $32\%$, where these savings are with respect to the case where {\it no energy management} is performed, i.e., the network is dimensioned for maximum expected capacity (maximum value of $\theta(\zeta,\gamma,t), M$ = $27$ \acp{VM}, $\forall \, t$).  A peak can be observed in the performance of \mbox{DETA-R} between $\SI{4} {\hour}$ and $\SI{7} {\hour}$ where it approaches close to ENAAM. This is due to a decrease in the expected traffic load, which translates into low server utilization ($0 \leq \gamma \leq 0.5$) and high energy savings. Note that $\alpha=0$ leads to the highest energy savings, where the BS radio frontend is moved as often as possible into sleep mode and the minimum number $b$ of VMs is active in all time slots. While this may be meaningful as an energy consumption lower-bound, a more interesting choice is provided by $\alpha > 0$, where the local processing cost is also taken into account. 

In Fig.~\ref{fig:savings_0.5} ($\alpha=0.5$), the ENAAM scheme achieves \acp{ES} of about $56\%$ and \mbox{DETA-R} of $29\%$ on average. As expected, this shows a reduction in energy savings compared to when $\alpha = 0$ .This is due to the balance between the emphasis on energy savings and \ac{QoS}, i.e., locally computed tasks, within the remote site. The evolution of \acp{ES} with respect to $\alpha$ is presented in Fig.~\ref{fig:alpha_val}. As expected, a drop in energy savings is observed when \ac{QoS} is prioritized, i.e., $\alpha \to 1$, as in this case the BS energy consumption is no longer considered.
 
\begin{figure}[t]
	\centering
	\includegraphics[width =\columnwidth]{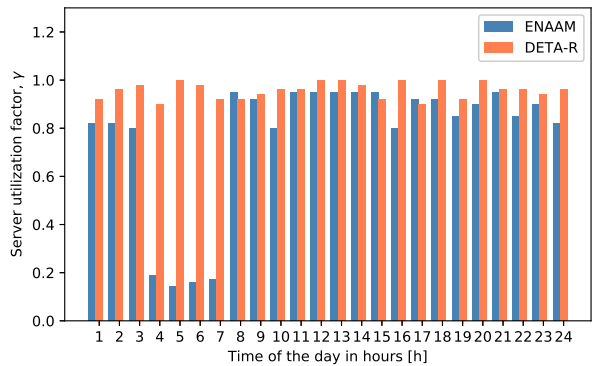}
	\caption{\ac{MEC} server utilization ($\alpha = 0.5$).}
	\label{fig:mec_2}
\end{figure}

Finally, Fig.~\ref{fig:mec_2} shows the MEC server utilization over time, i.e., the selected control ${\gamma}(t)$. For $\alpha = 0.5$, the server utilization is about $76\%$ for ENAAM and $95\%$ for \mbox{DETA-R} on average. A low server utilization can be observed in the performance of ENAAM between $\SI{4} {\hour}$ and $\SI{7} {\hour}$ due to an expected low traffic load in the system. This indicates that ENAAM has load adaptation capabilities, which are much desirable and lead to substantial energy savings (see again Fig.~\ref{fig:savings_0.5} between $\SI{4} {\hour}$ and $\SI{7} {\hour}$). 


\section{Conclusions}
\label{sec:concl}

In this paper, we have envisioned a \mbox{renewable-powered} remote site for extending network coverage and promote energy \mbox{self-sustainability} within mobile networks. The BS at the remote site is endowed with computation capabilities for guaranteeing low latency to mobile users, offloading their workloads. The combination of the energy saving methods, namely, \ac{BS} sleep modes and \ac{VM} \mbox{soft-scaling}, for the remote site helps to reduce its energy consumption. An edge energy management algorithm based on forecasting, control theory and heuristics, is proposed with the objective of saving energy within the remote base station, possibly making the BS system \mbox{self-sustainable}. Numerical results, obtained with \mbox{real-world} energy and traffic traces, demonstrate that the proposed algorithm achieves energy savings between $56\%$ and $66\%$ on average, with respect to the case where no energy management techniques are applied, and to hold the server utilization between $30\%$ and $96\%$ over time, with an average of $75\%$.

\section{Acknowledgements}

This work has received funding from the European Union's Horizon 2020 research and innovation programme under the Marie Sklodowska-Curie grant agreement No. 675891 \mbox{(SCAVENGE)}.

\bibliographystyle{IEEEtran}
\scriptsize
\bibliography{biblio}
\end{document}